\documentclass[aps,twocolumn,pra]{revtex4}
\usepackage{graphicx}

\newcommand\ro{\hat\rho}
\newcommand\Ho{\hat H}
\newcommand\Qo{\hat Q}
\newcommand\fo{\hat f}
\newcommand\Schr{Schr\"odinger}

\begin{document}
% Full title of the paper (Capitalized)
\title{Fundamental Irreversibility: Planckian or \Schr-Newton?}

\author{Lajos Di\'osi}

\affiliation{Wigner Research Centre for Physics, H-1525 Budapest 114, P.O.Box 49, Hungary; diosi.lajos@wigner.mta.hu}

\begin{abstract}
The inception of a universal gravity-related irreversibility took place originally in quantum cosmology.
The ultimate reason of universal irreversibility is thought to come from
black holes close to the Planck scale. Completely different instances of
irreversibilities are quantum state reductions unrelated to gravity or relativity
but related to measurement devices. However, an intricate relationship between Newton gravity and 
quantized matter might result in fundamental and
spontaneous quantum state reduction --- in the non-relativistic \Schr-Newton
context. The above two concepts of fundamental
irreversibility emerged and evolved with few or even no interactions.
The purpose here is to draw a parallel between the two approaches first,
and to ask rather than answer the question: can both the Planckian and the \Schr-Newton
indeterminacies/irreversibilities be two faces of the same universe.
A related personal note of the author's 1986 meeting with Aharonov and Bohm is appended. 
\end{abstract}

\maketitle
%%%%%%%%%%%%%%%%%%%%%%%%%%%%%%%%%%%%%%%%%%
\section{Introduction}
Standard micro-dynamical equations, whether classical or quantum, are 
deterministic and reversible. They can, nonetheless, encode various options 
of irreversibility even at the fundamental level. Here I am going to discuss two
separate concepts of fundamental irreversibility,  which are 
quite certain to overlap on the long run. The first option concerns
space-time (gravity), it is relativistic, hallmarked by mainstream
cosmologists and field theorists (including immortal ones).
The second option roots in the explicit irreversibility of 
von Neumann measurement in non-relativistic quantum mechanics, 
its story is perhaps more diffusive than the first one's.  
The standard and linear  story of Planck scale irreversibility is recapitulated
in Sec. \ref{Planck}. I choose a personal account for the parallel story of 
the conjectured Newton-gravity-related non-relativistic irreversibility of
macroscopic quantum mechanics  in Sec. \ref{SN}. I stop both stories
with the 1980's when the same structure of heuristic master equations became
proposed for the two options of fundamental irreversible dynamics --- 
with different interpretations and regimes of significance, of course. 
Towards their reconciliation, Sec. \ref{P_or_SN} offers  some thoughts with 
the open end. 

%%%%%%%%%%%%%%%%%%%%%%%%%%%%%%%%%%%%%%%%%%
\section{Irreversibility at Planck Scale}
\label{Planck}
At the dawn of quantum-gravity research, Bronstein 
\cite{Bron36a,Bron36b,Gor05} discovered by heuristic calculations that the 
precise structure of space-time, contrary to the precise structure of electromagnetism, 
is unattainable if we rely on quantized motion of test bodies. The coming decades
brought up stronger and famous arguments for space-time unsharpness, 
unpredictability, its role in universal loss of information, of quantum coherence,
and of microscopic reversibility in general. 
Wheeler \cite{Whe62} found that smooth space-time changes into a foamy structure 
of topological fluctuations at the Planck scale. Bekenstein \cite{Bek72} gave the first 
exact quantitative proposal toward fundamental irreversibility, claiming
black holes have entropy:
\begin{equation}
S=\frac{k_B}{4\ell_{Pl}^2}\times(\mbox{black hole surface area}),
\end{equation}
where $k_B$ is Boltzmann's constant, $\ell_{Pl}$ is the Planck length. 
This was confirmed by Hawking \cite{Haw75} 
who showed that black holes emit the corresponding thermal radiation indeed.
Only a little later, he summarized the situation by stating the unpredictability
of quantum-gravity at the Planck scale, leading him to propose that quantum
field theory is fundamentally irreversible. Accordingly, the unitary scattering operator
$\hat S$ should be replaced  
by the more general superscattering operator $\$$ acting
on the initial density operator $\ro_{in}$ instead of the initial state vector:
\begin{equation}
\ro_{out}=\$ \ro_{in}\neq {\hat S}\ro_{in}{\hat S}^\dagger.
\end{equation}
To resolve the detailed irreversible (non-unitary) dynamics beyond Hawking's
superscattering, Ellis et al. \cite{Elletal84} proposed a simple quantum-kinetic (master) equation,
which Banks, Susskind and Peskin \cite{BanSusPes84} generalized as follows: 
\begin{equation}
\frac{d\ro}{dt}=-\frac{i}{\hbar}[\Ho,\ro]-\frac{1}{2\hbar^2}\int\!\!\int [\Qo(x),[\Qo(y),\ro~]]h(x-y)d^3xd^3y,
\end{equation}
where $\Ho$ is the Hamiltonian, $\Qo(x)$ is a certain quantum field, 
and $h(x-y)$ is a positive symmetric kernel. The transparent structure allowed the authors to point
out a substantial difficulty: non-conservation of energy-momentum.

%%%%%%%%%%%%%%%%%%%%%%%%%%%%%%%%%%%%%%%%%%
\section{Irreversibility  in the \Schr-Newton context}
\label{SN}
In the early 1970's, being a student fascinated already by the quantum theory,
I missed a dynamical formalism of state vector collapse from it. Weren't I be a student, 
were I aware of the related literature, I would have read the phenomenological model 
by Bohm and Bub \cite{BohBub66}. 
But I was not aware of it, started to think on my own. Open the textbook,
you read the expansion of the time-dependent state vector $|t\rangle$ in terms of the energy
eigenstates $|n\rangle$ of eigenvalues $E_n$, resp. But I wrote it with a little modification:   
\begin{equation}
|t\rangle = \sum_n c_n\exp\left(-\frac{i}{\hbar}E_n (1+\delta) t \right)|n\rangle,
\end{equation}
because I observed that allowing a small randomness $\delta$ of the time flow,
the average density matrix becomes gradually diagonal in the energy basis:
\begin{equation}
\overline{|t\rangle\langle t|} \longrightarrow \sum_n |c_n|^2 |n\rangle\langle n|,
\end{equation}
exactly as
if someone measured the energy. I got a prototype dynamical model 
of non-selective von Neumann measurements. A question remained
to answer: where does randomness of time come from? The hint should have 
come from the sadly forgotten Bronstein \cite{Bron36a,Bron36b,Gor05} basically, but it
came occasionally from K\'arolyh\'azy after he gave department seminars in 1973 
on his earlier work \cite{Kar66} where he used a Planck scale uncertainty
of classical space-time and a very vague model of massive body's state
vector collapse based upon it. Just I had to do experimental particle physics for a decade. 

When back to theory, I showed \cite{Dio84} that the Newtonian
limit of standard reversible semiclassical  gravity, the so-called \Schr-Newton
equation \cite{Pen96}, obtains sensible solitonic wave functions for
the massive (e.g.: nano-) objects' center-of-mass.  It determined my way, as
to put {\em non-relativistic} flesh on the toy dynamics (4-5) of state vector reduction. 
The uncertainty $\delta$ of time flow should come from
of metric tensor element $g_{00}$, which is the Newton potential $\phi$ in fact.
The unpredictability $\delta\phi$ of the Newton potential should depend on $G$ 
and $\hbar$, but not on $c$. The choice was the following  spatially correlated white-noise:
\begin{equation}
\overline{\delta\phi(x,t)\delta\phi(y,s)}
=\frac{\hbar G}{\vert x-y\vert}\delta(t-s).
\end{equation}
The random part of the Newton potential couples to the mass density operator $\fo(x)$
via the interaction $\int\phi(x,t)\fo(x)d^3x$, yielding the following master equation
for the density operator:
\begin{equation}
\frac{d\ro}{dt}=-\frac{i}{\hbar}[\Ho,\ro]-\frac{G}{2\hbar}\!\!\int\!\!\!\int [\fo(x),[\fo(y),\ro~]]\frac{1}{\vert x-y\vert}d^3xd^3y.
\end{equation}
This dynamics is mimicking the (non-selective) von Neumann measurement
of massive object's positions, it predicts the {\em spontaneous reduction} (decay) 
of \Schr~cat states (see same result in \cite{Pen96} by Penrose).

Before journal publication \cite{Dio87}, I showed this result to Yakir Aharonov 
(read Sec. \ref{AB}).  He warned me of the energy-momentum non-conservation. I took it with surprise
because I did not read \cite{BanSusPes84}. 
%%%%%%%%%%%%%%%%%%%%%%%%%%%%%%%%%%%%%%%%%%
\section{Planck scale or \Schr-Newton context?}
\label{P_or_SN}
Irreversibility at the Planck scale seems
plausible within standard physics because of evaporating black holes 
(Sec. \ref{Planck}). The non-relativistic \Schr-Newton irreversibility   
(Sec. \ref{SN}) is a conjecture although its derivation is not
seriously more heuristic than the Planckian's.
For both options, the same structure of master equations were
proposed to encode the irreversible dynamics of the density operator.
Planck scale irreversibilities from eq. (3) become significant
for certain fundamental elementary particles. Contrary to that,
eq. (7) predicts irreversibility for massive non-relativistic objects
in the \Schr-Newton context.
Whether the two underlying concepts are compatible at all, it is not known.
Whether or not the Newtonian unpredictabilities/fluctuations are the
non-relativistic limit of the Planckian's? That is hard to answer.

Let me mention, nonetheless, two examples where relativistic 
phenomenologies, different from the line of Sec. \ref{Planck},
turned out to reduce to the \Schr-Newton uncertainty  (6) 
non-relativistically. Unruh \cite{Unr84} proposed a possible 
uncertainty relation between the metric and Einstein tensors, resp.
In the Newtonian limit, speed of light $c$ cancels and we are left
just with the white-noise uncertainties (6), as pointed out in \cite{Dio87}. 
Penrose discussed the fundamental conflict between general relativity 
and quantization. To resolve it heuristically at least, he also found the necessity of 
space-time's fundamental unsharpness, guessed it non-relativistically
and concluded to what was equivalent with expression (6) up to a factor 2 
(which discrepancy has recently been resolved by \cite{TilDio17}). 

Against questioning a possible transmutation
of Planck scale uncertainties into the non-relativistic \Schr-Newton regime,
I have an elementary argument .  Consider the \Schr-equation for the
center-of-mass of a big body like $M=1kg$, with velocity $1km/s$ which is fairly non-relativistic.
Calculate the de Broglie wave length: $\lambda=(2\pi\hbar/mv)=4.16\times10^{-36}m$.
This is smaller than the Planck length $\ell_{Pl}=1.62\times10^{-35}m$ by about one order
of magnitude. Since standard physics breaks down anyway at the Planck scale,
we can no longer trust in the \Schr~equation for the motion of our massive
non-relativistic body. Planck scale space-time uncertainties have thus
flown down into uncertainties in the \Schr~dynamics of non-relativistic massive
bodies. So far so good. But shall $c$ cancel so that we get the
effective \Schr-Newton uncertainty (6-7) and the corresponding spontaneous
reduction for massive objects \cite{Dio87,Pen96}? 

%%%%%%%%%%%%%%%%%%%%%%%%%%%%%%%%%%%%%%%%%%
\section{Concluding remarks}
Two independent theories of relativistic and non-relativistic
fundamental irreversibility, resp., both related to the conflict between gravity and quantization, 
are in the scope of this work. One was conceived and would be
relevant in cosmology. The other one was born from the quantum measurement
problem and would modify the quantum mechanics of massive bodies even in the lab.
Their conceptions have been outlined in Secs. \ref{Planck} and \ref{SN}, respectively, 
including their basics without the details and later developments.  
Such restricted presentation sufficed to expose the issue 
put in the center of this work in Sec. \ref{P_or_SN}: what is the relationship between the Planckian
and the \Schr-Newton unpredictability of our space-time? 
The answer remains missing,  but our purpose has been to urge it.
In particular, we pointed out  that Planckian unpredictability survives
non-relativistically --- for massive macroscopic quantized degrees of freedom. 

%\appendixtitles{no} 
\appendix
\section{}
\label{AB}
\begin{figure}[h]
\centering
\includegraphics[width=8.5 cm]{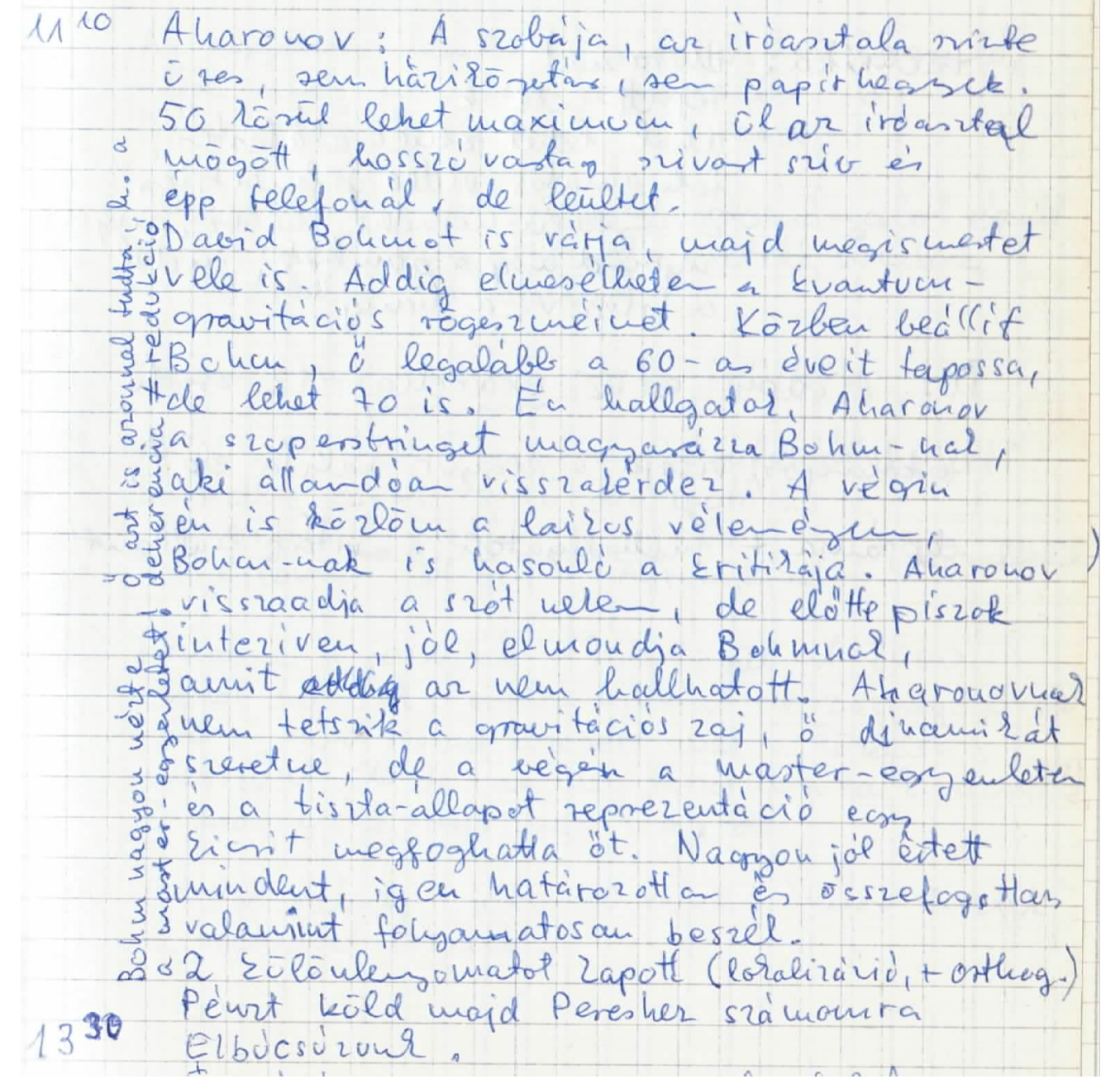}
\caption{Author's diary, page from March 18, 1986.}
\end{figure}   
It was the courtesy of Asher Peres who asked Yakir to receive the unknown
theorist from Hungary.  Below is the translation of my notes (Fig. A1).
\begin{quote}
11$^{10}$ Aharonov: His office and desk are almost empty, no personal library, no
paper piles. Maximum 50 or so, sits behind the desk, smokes long fat cigar
and just phones, but makes me seated.

Awaiting for David Bohm as well, shall introduce me to him as well. Until that,
I can unfold my quantum-gravity id\'ee fix. Meanwhile David Bohm arrives,
he is at least in his 60's, but can be 70. I'm listening, Aharonov
is explaining the superstring to Bohm who is repeatedly asking. Finally
I also communicate my layman's views, Bohm's criticism is also akin.
Aharonov returns the word to me, but first tells Bohm hellish intensively
what  he could not have heard. Aharonov dislikes the gravitational noise,
he'd prefer dynamics, but at the end my master equation and the
pure state representation may have caught him a bit. He understood
everything very well, his talking is really firm and organized, also steady.

He got two offprints (localization + orthog.)

Shall send Peres money for me.

13$^{30}$ We say good bye.

{\em Left margin:} Bohm looked at the master equation strongly!
Immediately he knew also that decoherence$\neq$reduction.  
\end{quote}

%%%%%%%%%%%%%%%%%%%%%%%%%%%%%%%%%%%%%%%%%%
\acknowledgments{The Fetzer Franklin Fund is acknowledged for its generous covering 
my costs to attend the Emergent Quantum Mechanics 2017 conference 
(26-28 October, University of London) and  to publish this work in open access. 
I thank the Hungarian Scientific Research Fund under Grant No. 124351
and the EU COST Action CA15220 for support.}

%%%%%%%%%%%%%%%%%%%%%%%%%%%%%%%%%%%%%%%%%%
%\conflictsofinterest{The author declares no conflict of interests.}

%=====================================
% References, variant A: internal bibliography
%=====================================
%\reftitle{References}


\begin{thebibliography}{99}
\bibitem{Bron36a} Bronstein, M. Quantentheorie schwacher Gravitationsfelder. 
{\em Phys. Z. Sowjetunion} {\bf 1936}, {\em 9}, 140-157.
\bibitem{Bron36b} Bronstein, M.P. Kvantovanie gravitatsionnykh voln. 
{\em Zh. Eksp. Theor. Fiz.} {\bf 1936}, {\em 6}, 195-236.
\bibitem{Gor05} Gorelik, G.M. Matvei Bronstein and quantum gravity:
70th anniversary of teh unsolved problem.  {\em Usp. Fiz. Nauk} {\bf 2005}, {\em 48}, 1039-1053.
\bibitem{Whe62}  Wheeler, J.A. {\em Geometrodynamics}; Academic Press: New York, 1962.
\bibitem{Bek72} Bekenstein, J.D. Black holes and entropy. {\em Phys. Rev. D} {\bf 1973}, {\em 7}, 2333-2346. 
\bibitem{Haw75} Hawking, S.W. Particle creation by black holes. {\em Commun. Math. Phys.} {\bf 1975}, {\em 43}, 199-220.
\bibitem{Haw82} Hawking, S.W. The unpredictability of quantum gravity. {\em Commun. Math. Phys.} {\bf 1982}, {\em 87}, 395-415.
\bibitem{Elletal84} Ellis, J., Hagelin, S., Nanopoulos, D.V., and Srednicki, M.
Search for violations of quantum mechanics.  {\em Nucl. Phys. B} {\bf 1984}, {\em 241}, 381-405.
\bibitem{BanSusPes84} Banks, T., Susskind, L., and Peskin, M.E. Difficulties for the evolution of pure states into mixed states.
 {\em Nucl. Phys. B} {\bf 1984}, {\em 244}, 125-134.
\bibitem{BohBub66} Bohm, D., and Bub, J. A proposed solution of the measurement problem in quantum mechanics by a hidden variable theory.
 {\em Rev. Mod. Phys.} {\bf 1966}, {\em 38}, 453-469.
\bibitem{Kar66} Karolyhazy, F. Gravitation and quantum mechanics of macroscopic objects. {\em Nuovo Cim.} {\bf 1966}, {\em 42}, 390-402.
\bibitem{Dio84} Di\'osi, L. Gravitation and quantum-mechanical localization
of macro-objects. {\em Phys. Lett. A} {\bf 1984}, {\em 105}, 199-202.
\bibitem{Pen96} Penrose, R. On gravity's role in quantum state reduction.
{\em Gen. Relativ. Gravit.} {\bf 1996}, {\em 28}, 581-600.
\bibitem{Dio87} Di\'osi, L. A universal master equation for the gravitational
violation of quantum mechanics. {\em Phys. Lett. A} {\bf 1987}, {\em 120}, 377-381.
\bibitem{Unr84} Unruh, W.G. Steps towards a quantum theory of gravity.
In {\em Quantum Theory of Gravity};, edited by S. M.
Christensen, S.M., Ed.; Adam Hilger Ltd: Bristol, England, 1984; pp. 234-242.
\bibitem{TilDio17} Tilloy, A.; Di\'osi, L. Principle of least decoherence for Newtonian semi-classical gravity. 
{\em Phys. Rev. D} {\bf 2017}, {\em 96}, 104045-(6). 
\end{thebibliography}
\end{document}